\documentclass[prb,twocolumn,showpacs,preprintnumbers,amsmath,amssymb,superscriptaddress]{revtex4}
\usepackage{epsf}
\usepackage{graphicx}
\usepackage{bm}
\usepackage{amsmath}
\usepackage{color}
\usepackage{notes2bib}
\usepackage{epstopdf}

\bibnotesetup{
note-name = , use-sort-key = false }
\begin{document}

\title{Scattering on magnetic skyrmion in the non-adiabatic approximation}

\author{K.~S.~Denisov}
\email{denisokonstantin@gmail.com} \affiliation{Ioffe Institute, 194021 St.Petersburg, Russia}
\affiliation{Lappeenranta University of Technology, FI-53851 Lappeenranta, Finland}
\author{I.~V.~Rozhansky}
\affiliation{Ioffe Institute, 194021
St.Petersburg, Russia} \affiliation{Lappeenranta University of
Technology, FI-53851 Lappeenranta, Finland}
\affiliation{Peter the Great Saint-Petersburg Polytechnic University, 195251 St. Petersburg, Russia}
\author{N.~S.~Averkiev}
\affiliation{Ioffe Institute, 194021 St.Petersburg, Russia}
\author{E.~L\"ahderanta}
\affiliation{Lappeenranta University of Technology, FI-53851 Lappeenranta, Finland}

\begin{abstract}
We present a theory of electron scattring on a magnetic skyrmion for the case when exchange interaction is moderate so that adiabatic approximation and Berry phase approach is not applicable. The theory explains the appearance of a topological Hall current in the systems with magnetic skyrmions, of a special importance its applicability to dilute magnetic semiconductors with a weak exchange interaction.
\end{abstract}


\date{\today}

\maketitle

The last decade has been marked
by the discovering of numerous magnetic systems with exotic magnetic order: the steady magnetic skyrmions forming magnetic structures of different shapes \cite{BraunNano, NagaosaNature}, Kagome lattices \cite{KagomeLat1} or individual skyrmions located apart from each other \cite{Express}. Skyrmion \cite{Skyrme} is a special vortex configuration of spins. The skyrmion is distinguished from a
various rotational fluctuations of magnetization by possessing nontrivial topological invariant $Q$ called winding number (\ref{skyrm}). Since $Q\neq 0$, the magnetic skyrmion exhibits a number of properties, specific only for the topologically non-trivial structure. Moreover, the topological protection makes the skyrmion structure resistant to a smooth disorder, therefore a skyrmion is considered as a promising candidate for information storage unit \cite{SkApplic-1, SkApplic-2}. This has drawn a special attention to systems with a small number of magnetic skyrmions \cite{Romming2013, Express}.
The key tool for probing of a single magnetic skyrmion is the topological Hall effect (THE) \cite{Ye1999, BrunoDugaev}, the state of each magnetic skyrmion can be read out by measuring the Hall-voltage.  The topological Hall response with the discrete features indicating the flipping magnetization of every single skyrmion was experimentally demonstrated \cite{DiscretHall}. However, extracting THE contribution from the total Hall response meets certain difficulties  \cite{THE_Li, DiscretHall} originating from the lack of theoretical description of THE in various regimes.
\normalcolor

Unlike other mechanisms of the anomalous Hall effect (AHE)\cite{AHE-Sinova} THE is not based on spin-orbit interaction of the free charge carriers,
 it originates from the nontrivial spatial configuration of magnetization\cite{Ye1999}.
Phenomenologically the skyrmion is often regardered as a source of effective magnetic field acting on electrons. However, theoretically, this approach is justified only in adiabatic (Berry phase) approximation\cite{Ye1999,Lyana-Geller1} valid in systems with strong exchange interaction.
The adiabatic approach is valid when the oscillation frequency $\omega_{ex}$ between two itinerant electron spin sublevels split by the exchange interaction with magnetic centers exceeds the inverse time $\tau_{sk}^{-1}$ of electron flight through the skyrmion core.
In the adiabatic regime there are no quantum transitions between spin subleves and the axis of electron spin quantization moves adiabatically following the direction of the local magnetization and acquires a real-space Berry phase \cite{AronzonRozh, Berry1} which can be treated as effective magnetic field.
Although the adiabatic approximation is commonly used to treat
THE, it is not applicable for the important case
 of relatively weak exchange interaction  $\omega_{ex} \tau_{sk} \le 1 $. This case is related to the spin-flip scattering of electron on magnetic impurities and it is typical for some spin-glass systems \cite{Tatara, Nakazawa2014} and dilute magnetic semiconductors (DMS)\cite{BrunoDugaev}.
 Nowadays DMSs are considered as very promising materials
 for semiconductor spintronics as they provide various options for control and detection of electron spin polarization within the semiconductor element base. An experimental manifestation of THE
 in DMS has been also reported recently \cite{AronzonRozh}.
Therefore, a valid theoretical description of THE for the systems with weak exchange interaction is highly necessary.
Our present work fills this gap in the theory of THE by
\normalcolor
 considering the scattering on skyrmion in non-adiabatic regime with account for spin-flip transitions.
\normalcolor
Our main finding is that an electron scattering on skyrmion produces an electrical current in the same perpendicular direction regardless the electron spin.
Thus, the topological Hall current is not coupled to the carrier spin polarization.
This behavior completely differs from the other mechanisms of AHE, which contribute in electrical Hall response only if the carriers are spin polarized.
In addition to the magnetic field-like effect the skyrmion causes transverse spin-flip processes. Therefore the resulting effect on a carrier cannot be merely to an effective magnetic field, there is the additional spin-Hall contibution, which can be detected by spin accumulation at the edges of the sample.

\begin{figure}
	\centering\includegraphics[width=0.4\textwidth]{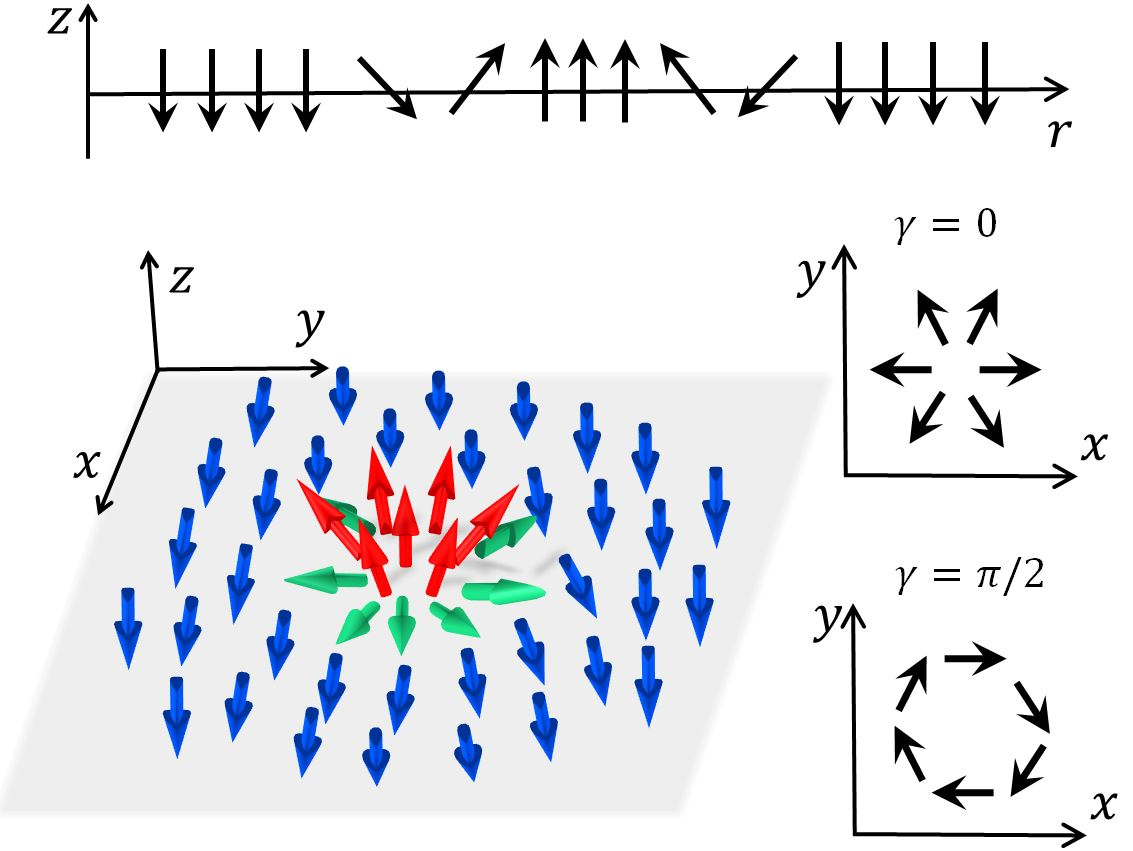}
	\caption{The profiles of the magnetic skyrmion with the winding number $Q=1$ and orientation $J_z = -1$, the helicity $\gamma=0$ and $\pi /2$.}
	\label{fig1}
\end{figure}

Our approach is
to calculate the cross-section of an itinerant electron on an ensemble of magnetic impurities forming a skyrmion. We do not account for spin-orbit interactions splitting the free electron states.
We are not interested in the effect of itinerant electron on the magnetic system  so we neglect the transitions between magnetic impurity spin states and describe the $i$-th magnetic impurity with a classical magnetic moment ${\bf I}_i = I {\bf m}_i$, aligned along the direction ${\bf m}_i$ determined by the skyrmion structure. 
We describe the spatial configuration of magnetic moments forming the skyrmion as:
\begin{equation}\label{skyrm}
{\bf m}({\bf r}) = \begin{pmatrix} \sin{\Theta(r)} \cos{\Phi(\varphi)}  \\ \sin{\Theta(r)} \sin{\Phi(\varphi)} \\  \cos{\Theta(r)} \end{pmatrix},
\end{equation}
where $\bf{r}=(r,\varphi)$ is the 2D radius vector,
$\Phi, \Theta$ are parameters of the skyrmion structure.
The skyrmion is a spatially localized twisted spin configuration.
Inside its core with size $a$ the spins are in opposite direction with the outside ones. This implies that $\Theta(r)$ must satisfy $\cos{\Theta(r \rightarrow 0)} \cos{ \Theta(r \rightarrow \infty)} <0$, which can be described for instance by  $\Theta(r) = \pi e^{-r/a}$. We will denote skyrmion orientation as $J_z = {\rm sign}( \cos{\Theta(r \rightarrow \infty)})$, there are two modifications of a skyrmion characterized by $J_z=\pm 1$. 
For a skyrmion with non-zero topological invariant (winding number) $Q$ the parametric function $\Phi(\varphi)$ (\ref{skyrm}) has the form
$\Phi = Q \varphi + \gamma$, where $Q$ is the winding number taking integer values and $\gamma$ is known as helicity. A skyrmion with $Q>0$ has full rotation symmetry whereas the one with $Q<0$ does not have it. The three parameters $Q$,$J_z$,$\gamma$ fully define a magnetic skyrmion. For example, a
skyrmion with  $Q=+1$, $J_z=-1$, and $\gamma=0, \pi/2$ is shown in Fig.\ref{fig1}.
The magnetic impurities forming a skyrmion usually also form a magnetic order outside its core. Since we neglect spin-orbit splitting of the free carriers spectra, the magnetic impurities outside $r>a$ don't contribute to AHE so will not account for those while considering the Hall current

We consider the contact exchange interaction between an electron and magnetic impurities in the form:
\begin{equation}
\hat{V}_{ex} = - \alpha_e \sum_{i}{\delta({\bf r}-{\bf R}_i)  \hat{{\bf s}} {\bf I}_i},
\end{equation}
where $\alpha_{e}$ is the exchange constant (we assume ferromagnetic exchange interaction, i.e. $\alpha_e > 0$), the sum is over all magnetic centers, ${\bf R}_i$ is the position of $i$-th magnetic center,
$\hat{{\bf s}}$ is the electron spin operator. The magnetic impurity spin ${\bf I}_i$ acts on an electron as a magnetic field and allows for spin-flip processes (if not aligned with the electron spin).
Let us consider the 2D elastic scattering of an electron on the skyrmion core. The initial state is a plane-wave characterized with momentum $\textbf{p}$ and spin index $\alpha$, $|\bf{p},\alpha \rangle$, the final state is $|\bf{p'},\beta \rangle$.
The differential scattering cross-section is given by:
\begin{equation}\label{eq_cross-1}
d\sigma_{\alpha \beta} (\theta) = \frac{m^2}{2 \pi \hbar^3 p} \left| F_{\alpha \beta}({\bf p}, {\bf p}') \right|^2 d\theta,
\end{equation}
where $F_{\alpha \beta}({\bf p}, {\bf p'})$ is the scattering amplitude, $\theta$ is the scattering angle, $m$ is the electron effective mass, $p = |\textbf{p}|=|\textbf{p}'|$.
$F_{\alpha \beta}({\bf p}, {\bf p}')$ is to be calculated by means of  perturbation theory with respect to the exchange constant $\alpha_e$.
The spatial correlation of the magnetic moments forming the skyrmion manifests itself in the second order of perturbation theory, which takes into account the particle propagation between two subsequent scatterers.
The scattering amplitude in this approximation is given by:
\begin{multline}
\label{eqF}
F_{\alpha \beta}({\bf p}, {\bf p}') =
V_{\alpha \beta}({\bf p}' - {\bf p}) +
\\ + \int V_{\alpha \gamma}({\bf p}'-{\bf q})
G_{\gamma \delta}^0({\bf q},\varepsilon) V_{\delta \beta}({\bf q}-{\bf p}) \frac{d{\bf q}}{(2\pi \hbar)^2},
\end{multline}
where $G_{\alpha \beta}^0({\bf q},\varepsilon) = \delta_{\alpha \beta}\left( \varepsilon - q^2/2m - i0 \right)^{-1}$ is the free electron Green's function,  $\varepsilon = p^2 / 2m$, $V_{\alpha \beta}({\bf q})$ is the matrix element of the perturbation:
\begin{equation}
V_{\alpha \beta}({\bf q}) = - \frac{\alpha_e I}{2} \sum_{i}{ e^{-i{\bf q}{\bf R}_i/\hbar}
	\left( {\bf m}_i {\bf \sigma}\right)_{\alpha \beta} }.
\end{equation}
We consider the scattering on a symmetrical skyrmion with positive $Q$. In this case the differential cross section depends only on the scattering angle $\theta$.
Taking modulus squared of $F_{\alpha \beta}$ given by (\ref{eqF}) to be substituted into (\ref{eq_cross-1}) one gets terms with different powers of $\alpha_e$. The terms of the order of $\alpha_e^2$ are the Born approximation and describe the uncorrelated  scattering on individual centers.
To get the effect of the non-complanar skyrmion spin structure at least three impurity spins are to be considered.
Such a three-spin interaction is the interference of first and second-order processes, the corresponding terms in the differential cross section appear in the order of $\alpha_e^3$.
Analysis of the $\alpha_e^3$ terms reveals two distinct contribution to the differential cross section.
The first one denoted as $d\sigma_{\alpha \beta}^A$ (``asymmetrical'' ) is
specific for the non-complanar spatial configurations of magnetization, turns to zero for a configuration with all spins aligned.
In the order of $\alpha_e^3$ there is also a contribution described by scalar product of electron and impurity spins, which also exists for a trivial non-skyrmion case of all impurity spins aligned.
All such terms up to the order of $\alpha_e^3$ are absorbed into $d\sigma_{\alpha \beta}^S$ (``symmetrical''), so the differential cross section is the sum of the two contributions:
\begin{equation}
d\sigma_{\alpha \beta}(\theta) = d\sigma_{\alpha \beta}^{S}(\theta) + d\sigma_{\alpha \beta}^{A}(\theta).
\end{equation}
Although $d\sigma_{\alpha \beta}^S$ contribution obviously dominates by means of total cross section (as it contains lower order $\alpha_e^2$), it 
does not provide the asymmetry of the scattering in perpendicular direction required for the Hall response, thus in our further considerations of THE we focus only on  $d\sigma_{\alpha \beta}^A$ term.

The explicit expression for  $\Sigma_{\alpha \beta} \equiv d\sigma_{\alpha \beta}^A / d\theta $ is found to be:
\begin{equation}\label{simH}
\Sigma_{\alpha \beta} = A {\rm Im} \left[ \int d{\bf r}_1 d{\bf r}_2 G_0(r_{12},\varepsilon) e^{-i {\bf p}_2 {\bf r}_1+i {\bf p}_1 {\bf r}_2}
\hat{K}_{\alpha \beta}({\bf r}_1, {\bf r}_2) \right],
\end{equation}
where $A = m^2 (\alpha_e I n_m)^3 /8 \pi \hbar^3 p$, $n_m$ is the sheet density of the magnetic centers, $ G_0(r_{12},\varepsilon)$ is the Green's function in the coordinate - energy representation,
\[
\hat{K}({\bf r}_1, {\bf r}_2) = \hat{I} g_z({\bf r}_1, {\bf r}_2) + \hat{\sigma}_x (g_x({\bf r}_1, {\bf r}_2)+g_y({\bf r}_1, {\bf r}_2))
\]
\begin{equation}\label{g}
g_i({\bf r}_1, {\bf r}_2) = \int d{\bf r}  \left[{\bf m}({\bf r+r}_1) \times  {\bf m}({\bf r+r}_2)  \right]_i {\bf m}_i({\bf r})
\end{equation}
$\hat{\sigma}_x$ is the Pauli matrix, $\hat{I}$ is the unity matrix, $g_i({\bf r}_1, {\bf r}_2)$ is the spin-chirality correlation function of the magnetization, $i={x,y,z}$. It is the function $g_i$ that describes the geometrical properties of the magnetic skyrmion.
It contains vector product of magnetic moments (\ref{g}) and turns to zero for the topologically trivial case of aligned spins.

It appears that the contribution to the cross section $\Sigma_{\alpha \beta}$ (\ref{simH}) favours asymmetrical scattering. Indeed, because of $g_i({\bf r}_1, {\bf r}_2) = - g_i({\bf r}_2, {\bf r}_1)$ we conclude, that $\Sigma_{\alpha \beta}$ is odd with respect to the scattering angle: $\Sigma_{\alpha \beta}(\theta) = - \Sigma_{\alpha \beta}(-\theta)$. This leads to
different scattering rates for scattering to the upper and lower half-planes.
This other important properties following from (\ref{simH}) are:
\begin{equation}\label{eqSym}
\Sigma_{\uparrow \uparrow}(\theta) = \Sigma_{\downarrow \downarrow}(\theta) \,\,\,\,\,\,\,\,\,\,\,
\Sigma_{\uparrow \downarrow}(\theta) = \Sigma_{\downarrow \uparrow}(\theta).
\end{equation}
So, regardless of its spin an electron is mostly scattered to a preferred perpendicular direction.
From this point of view, the skyrmion effect on the charge transport is similar to an effective magnetic field acting on a spinless particle
with the field direction determined by the skyrmion orientation $J_z$.
This is consistent with the adiabatic approximation description.
Besides the orbital transverse effect, the magnetic skyrmion also causes spin conversion in the perpendicular direction described by $\Sigma_{\uparrow\downarrow}$. This is the topological spin Hall effect (TSHE) which is not described by the adiabatic approximation.

\begin{figure}
		\centering\includegraphics[width=0.4\textwidth]{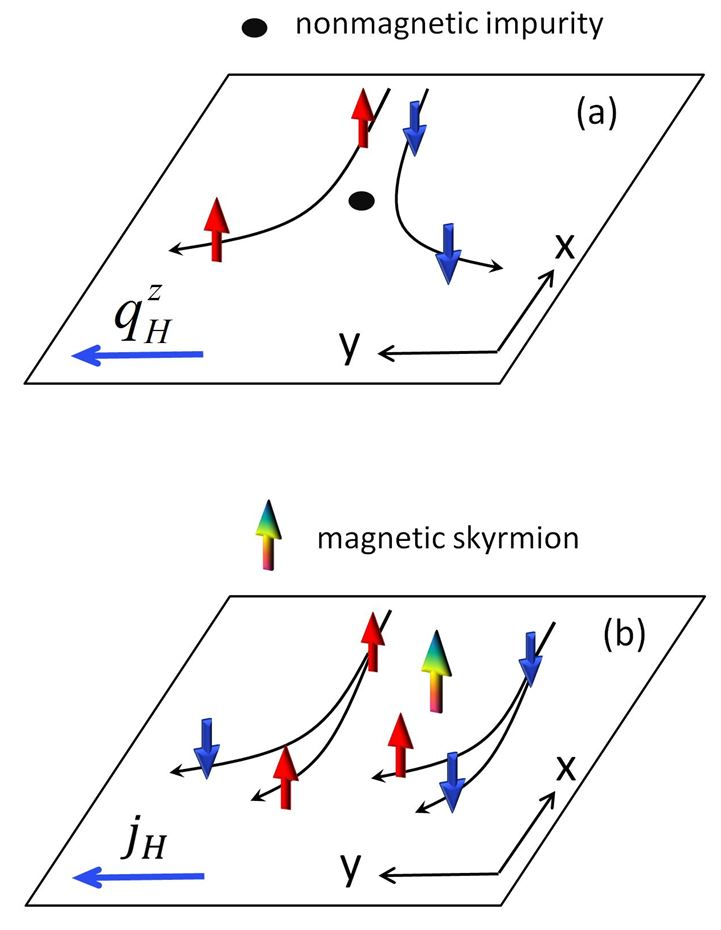}
		\caption{Different mechanisms of the $\rho_{yx}$. (a) AHE - skew-scattering. (b) THE - scattering on magnetic skyrmion}
		\label{fig2}
	\end{figure}

Let us discuss the phenomenology of the Hall effect for the media containing magnetic skyrmions.
An external electric field $E_x$ applied along along $x$-axis produces an electrical current $j_x^{(0)}$, the perpendicular Hall current $j_y$ is given by $j_y = \rho_{yx} E_x$. The transverse resistivity $\rho_{yx}$ contains, in general, three contributions: $\rho_{yx} = \rho_{yx}^N +\rho_{yx}^A + \rho_{yx}^T$, where $\rho_{yx}^N$ is the ordinary Hall coefficient, $\rho_{yx}^A$ is the anomalous Hall effect contribution, $\rho_{yx}^T$ is due to the topological Hall effect. The normal Hall coefficient  $\rho_{yx}^N$
can usually be determined from resistance measurements in strong magnetic field.
The other two contributions $\rho_{yx}^A$ and $\rho_{yx}^T$ arise from the the internal magnetization. As our theory shows there is a fundamental difference between the two.
The microscopic mechanism underlying an anomalous Hall effect (AHE) is
separation of electrons with opposite spins by scattering in the opposite perpendicular directions Fig.\ref{fig2}(a). When the number of spin-up and spin-down electrons is different, the imbalance in the up and down scattering produces an electrical current in the perpendicular direction.
Thus, the observation of AHE essentially requires spin-polarization of the carriers in the sample.
On the contrary, the considered THE is driven by the skyrmion orientation $\rho_{yx}^T \sim J_z$ and it is irrelevant to the electron spin polarization.
  This opens a way to detect the THE experimentally.
   The systems with $\omega_{ex} \tau_{sk} \le 1$ have a long spin relaxation time (and large spin diffusion length) compared with strong ferromagnets, this makes it possible to inject the carriers with a certain spin polarization.
   The topological contribution would manifest itself by the suppressed  $\rho_{yx}$ dependence on spin polarization and, in particular, by the appearance of the Hall voltage even for non-polarized carriers.
The scattering on a magnetic skyrmion considered above also leads to the topological spin Hall effect (TSHE). The spin-flip terms $\Sigma_{\uparrow\downarrow}$, $\Sigma_{\downarrow\uparrow}$ 
	give rise to the perpendicular spin current $q_y \sim J_z S_z$.	This TSHE is in contrast with other known mechansisms of spin-Hall effect where transverse spin current exists independently of the carrier spin polarization.

The expression (\ref{simH}) can be explicitly calculated for a simplified limiting case
when the electron wave-length $\lambda_F$ is small compared with the skyrmion size $a \gg \lambda_F$.
In this case it is reduced to:
\begin{equation}
\Sigma(\theta)=  \frac{(m \alpha_e  I n_m)^3 (\pi a)^2}{4 \hbar p^5}  J_z Q \sin{\theta}
\end{equation}
where $\Sigma(\theta) \equiv \Sigma_{\uparrow \uparrow} + \Sigma_{\downarrow \uparrow}  = \Sigma_{\downarrow \downarrow} + \Sigma_{\uparrow \downarrow}$ is
the cross section (\ref{simH}) summed over spin projections of the final state.

The non-adiabatic approximation considered in this paper is
of particular importance for systems with a weak exchange interaction.
Let us discuss one example of such system -- a II-VI semiconductor macrostructure with a quantum well (QW) doped with Mn or Fe atoms. These magnetic centers are isovalent and therefore electrically neutral in II-VI semiconductors so they interact with mobile charge carriers only by means of exchange interaction. Formation of skyrmions by an artificial triangular lattice of magnetic nano cylinders in a semiconductor QW was earlier proposed in \cite{BrunoDugaev}, 
We propose another mechanism of skyrmion formation in a thin semiconductor QW with spin-orbital interaction.
Magnetic skyrmion can be formed
on the basis of a bound magnetic polaron. An electron coupled to a donor impurity acquires a spin polarization in the presence of background magnetic impurities due to an effective exchange field (bound magnetic polaron).
In the presence of Rashba spin-orbit interaction\cite{Kavokin} this exchange field gives rise to the non-trivial internal spin configuration\cite{DenisovPolaron} and the magnetic polaron acquires non-zero winding number $Q$.
Regardless particular mechanism of a magnetic skyrmion or similar topological magnetic structures formation in a semiconductor QW, this system requires the  non-adiabatic approach. Indeed, the exchange splitting of an electron state is about $\Delta \sim 2$ meV for typical sheet density of the magnetic centers $n_m \approx 10^{12}$ cm$^{-2}$
, so the corresponding quantum transitions frequency is $\omega_s = \Delta/\hbar \approx2\cdot10^{12}$ s$^{-1}$
To estimate the Fermi velocity we take Fermi energy $E_F\approx 10$ meV and the effective mass in CdMnTe $m^{\ast}=0.11 m_0$, $m_0$ being the free electron mass, that makes
$v_F\approx2\cdot10^{7}$ cm/s.
With that it takes the time $\tau_{sk} = \approx2\cdot10^{-13}$ for the electron to pass the magnetic skyrmion core of a size of $a\approx30$ nm.
With $\omega_s \tau_{sk} \sim 0.5$ the non-adiabatic approximation is essential for this system.

Considering the AHE and THE in real samples one should along with the magnetic centers account for non-magnetic scatterers.
If the conductance of a sample is limited by the non-magnetic scatterers
their presence inside the skyrmion core can destroy the coherent scattering on magnetic centers\cite{Nakazawa2014, Tatara}.
However, for the dilute magnetic semiconductors and other systems with longitudinal conductivity limited by the magnetic impurities or
static potential disorder (with a spatial scale exceeding $a$) the suppression of THE is not expected.

 To summarize, we have described the topological Hall effect and spin Hall effect due to elastic scattering of itinerant carriers on a magnetic skyrmion in the non-adiabatic approximation $\omega_s \tau_{sk} \le 1$. We emphasize that unlike anomalous Hall effect the topological Hall electrical current does not depend on electron spin polarization. 
 
 This work has been carried out under the financial support of a
Grant from the Russian Science Foundation(Project no.14-12-
00255). 

\bibliography{Skyrmion}

\end{document}